\documentclass[fleqn, 10pt]{wlscirep}
\usepackage[utf8]{inputenc}
\usepackage[normalem]{ulem}

\title{Defining a historic football team: Using Network Science to analyze Guardiola's F.C. Barcelona}

\author[1,2,3 *]{J. M. Buld\'u}
\author[4]{J. Busquets}
\author[1,2]{I. Echegoyen}
\author[5]{F. Seirul.lo}

\affil[1]{Complex System Group \& GISC, Universidad Rey Juan Carlos, Madrid, Spain}
\affil[2]{Laboratory of Biological Networks, Center for Biomedical Technology, Universidad Polit\'ecnica de Madrid, Madrid, Spain}
\affil[3]{Institute of Unmanned System and Center for OPTical IMagery Analysis and Learning (OPTIMAL), Northwestern Polytechnical University, Xi'an 710072, China}
\affil[4]{E.S.A.D.E. Business School, Barcelona, Spain}
\affil[5]{Departamento de Metodolog\'ia, F.C. Barcelona, Barcelona, Spain}

\affil[*]{Corresponding author: jmbuldu@gmail.com}

\keywords{football, soccer, data analysis, network science, F.C. Barcelona, performance}

\begin{abstract}
The application of Network Science to social systems has introduced new methodologies to analyze classical
problems such as the emergence of epidemics, the arousal of cooperation between individuals or the propagation
of information along social networks. More recently, the organization of football teams and their performance have been unveiled using 
metrics coming from Network Science, where a team is considered as a complex network whose nodes (i.e., players) interact 
with the aim of overcoming the opponent network.  Here, we combine the use of different network metrics to extract the particular 
signature of the F.C. Barcelona coached by Guardiola, which has been considered one of the best teams along football history. We have first compared
the network organization of Guardiola's team with their opponents along one season of the Spanish national league, identifying those metrics 
with statistically significant differences and relating them with the Guardiola's game. Next, we have focused on the temporal nature of football passing networks 
and calculated the evolution of all network properties along a match, instead of considering their average. In this way, we are able to identify those network metrics
that enhance the probability of scoring/receiving a goal, showing that not all teams behave in the same way and how the organization Guardiola's F.C. Barcelona is different from the rest, including
its clustering coefficient, shortest-path length, largest eigenvalue of the adjacency matrix, algebraic connectivity and centrality distribution. 
\end{abstract}

\begin{document}
\flushbottom
\maketitle
\section*{Introduction}

Social systems have been one of the fields that has benefited the most from the wide variety of methodologies comprised under the umbrella of 
Network Science \cite{girvan2002,barabasi2002,eubank2004,vegaredondo2007,centola2010}.
Using such an approach, it is possible (i) to identify the most influential individuals of a social network \cite{katz1953,wasserman1994,newman2001,borgatti2005,newman2005,goldenberg2009}, 
(ii) to detect 
the existence of communities of people and the common interests that
tie them more tightly than individuals in other communities \cite{teitelbaum2008,fortunato2010,fortunato2016}, (iii) to explain the 
propagation of rumors/diseases \cite{pastor2001,newman2002,nekovee2007,balcan2009} or (iv) to analyze the bursting activity
of individuals when communicating with others \cite{barabasi2005}, just to cite a few examples.
Furthermore, the areas of application and systems under study are as diverse as (i) on-line social networks 
(e.g., Facebook or Twitter) \cite{ugander2011,goncalves2011,gonzalezbailon2011,grabowicz2012,weng2013,cheng2014}, (ii) interactions between companies 
and shareholders \cite{koenig2009,vitali2011}, (iii) crime networks \cite{ferrara2014}, (iv) collaborations 
between scientists \cite{newman2001,barabasi2002}, or (v) scaling laws in cities \cite{li2017}.

From the diversity of applications of Network Science, here we are concerned about the analysis of football matches and, specifically, the way players interact
with each other by passing the ball, ultimately creating what is known as a {\it football passing network}. 
Passing networks are constructed from the observation of the ball exchange between players, where network nodes (or vertices) are football players 
and links (or edges) account for the number of passes between any two players of a team. This way, we can construct football passing networks, weighted and unidirectional, which in turn are spatially embedded \cite{newman2010,ramos2018,buldu2018} (see Methods for an example about how passing networks are built).
The seminal paper by Gould and Gatrell \cite{gould1979}, published in the late seventies, introduced the concept of 
passing networks associated to a football match. However, it did not obtain the relevance it deserved, both in the scientific and sports communities. 
More than thirty years later, the work of Duch and collaborators \cite{duch2010} marked the start of a decade 
that is witnessing how the analysis of passing networks (by means of Network Science) is unveiling crucial information about the organization, evolution 
and performance of football teams and players \cite{buldu2018}. 

For example, inspecting the organization of passing networks, it is possible to detect recurring pass sequences and relate them to the playing style of a team \cite{gyarmati2016}. Passing networks, taken as whole, exhibit a small-world topology \cite{narizuka2014}, typically with high clustering coefficient (i.e., a tendency to create triangles of passes between three players) when compared to a random null model \cite{cotta2013}, and 
where the number of steps to go from one node to 
any other is much lower than the number of nodes of the network \cite{watts1998}. 
It is also possible to detect the existence of {\it motifs} \cite{milo2002}, consisting in the overabundance of certain kinds of passes between groups of three/four players \cite{gyarmati2014} or even  communities of players tightly connected between them \cite{clemente2015}. 
When the focus is placed at the player level, we can use network motifs to characterize the role of a player in a team or even to find players (in other teams) with similar features \cite{lopez2015}.
Furthermore, importance of players in a passing network can be quantified using the betweenness or closeness parameters, 
which show that passing networks are prone to find a balance between all players \cite{goncalves2017}.

Taking advantage of this new point of view that Network Science can give to the analysis of football datasets, we are going to analyze the particular organization of F.C. Barcelona (FCB) during the supervision of Pep Guardiola, considered as the referent team during the last decade \cite{whitehouse2014}.
Going back in time, modern football was invented in England and we can trace its rules back to 1863. In the beginning, team strategy consisted in moving forward the fastest and getting rid of the ball as soon as possible: having the ball, specially close to your goal, was seen as something ``dangerous". Teams were originally organized in a ``static" manner 
with clear and specialized roles of defenders, midfielders and forwards. In the 1950s, the Hungarian national team started to 
consider the ball as ‘not dangerous’ and organized the game around it. This led to a more dynamic approach in the 
1970s, giving birth to a new game system played by AFC Ajax, and the 
Dutch national team, which was called {\it ``total football"} ({\it totaalvoetbal} in Dutch). Rinus Michels and Johan Cruyff were responsible for this new style. Its development in F.C. Barcelona gradually 
appeared when Michels served as the club’s manager/coach (1971-1975 and 1976-1978) followed by others such as 
Johan Cruyff (1988-1996) and, definitively, Frank Rijkaard (2003-2008) and Pep Guardiola (2008-2012) \cite{caruncho2017,wilson2018}. The style of the 
Spanish national team (2008-2012) was also similarly influenced.
 
The tactical ability of Guardiola, which relied on a sophisticated combination of possession and pressing that, in turn, were synchronized to the positional play of the team, leaded to the most fruitful period of F.C. Barcelona, both in reputation and in the number of titles achieved, including $14$ titles during $4$ seasons.
In a more general framework, Guardiola was not the first coach who focused on pressing and possession or any of the other principles that, as he admitted, 
were extracted from the philosophy of his former coach Johan Cruyff \cite{wilson2018}. 

Despite there exists a vast literature about the particular features of Guardiola's teams \cite{balague2012,violan2014,perarnau2016}, quantitative analyses of their game style are still scarce.
With the aim of supporting the evidence with numbers, we are going to use Network Science to provide a different perspective of FCB style of playing, a perspective
focused on the organization of FCB passing networks and their differences with the rest of the teams paying in the Spanish national league.
We are going to focus on the season 2009/2010,
probably the most fruitful season of Guardiola's period, achieving the titles of six major 
competitions (Spain's Super Cup, UEFA Super Cup, FIFA Club World Cup, King's Cup, La Liga, and the UEFA Champions League). 
First, we will obtain the passing networks corresponding to the $380$ matches of ``La Liga" national league during the 2009/2010 season.
Next, we will analyze the differences between Guardiola's team 
and the rest of Spanish teams, identifying similarities and differences at the network parameters and linking them with the particularities of 
Guardiola's principles. At this point, we will discuss the influence of the temporal fluctuations of the network parameters along
a match and will propose a temporal analysis of passing networks. With this aim, we will introduce the concept of {\it 50-pass networks}
and recalculate all network parameters at different moments of the match, giving special attention to scored/received goals. When time is taken into account, our results show that (i) 
passing networks unveil additional information not contained in the average network and, in addition, (ii) temporal analysis highlights some of the particular features of Guardiola's game.

\begin{figure}[!ht]
 \centering
 \includegraphics[width=0.95\textwidth]{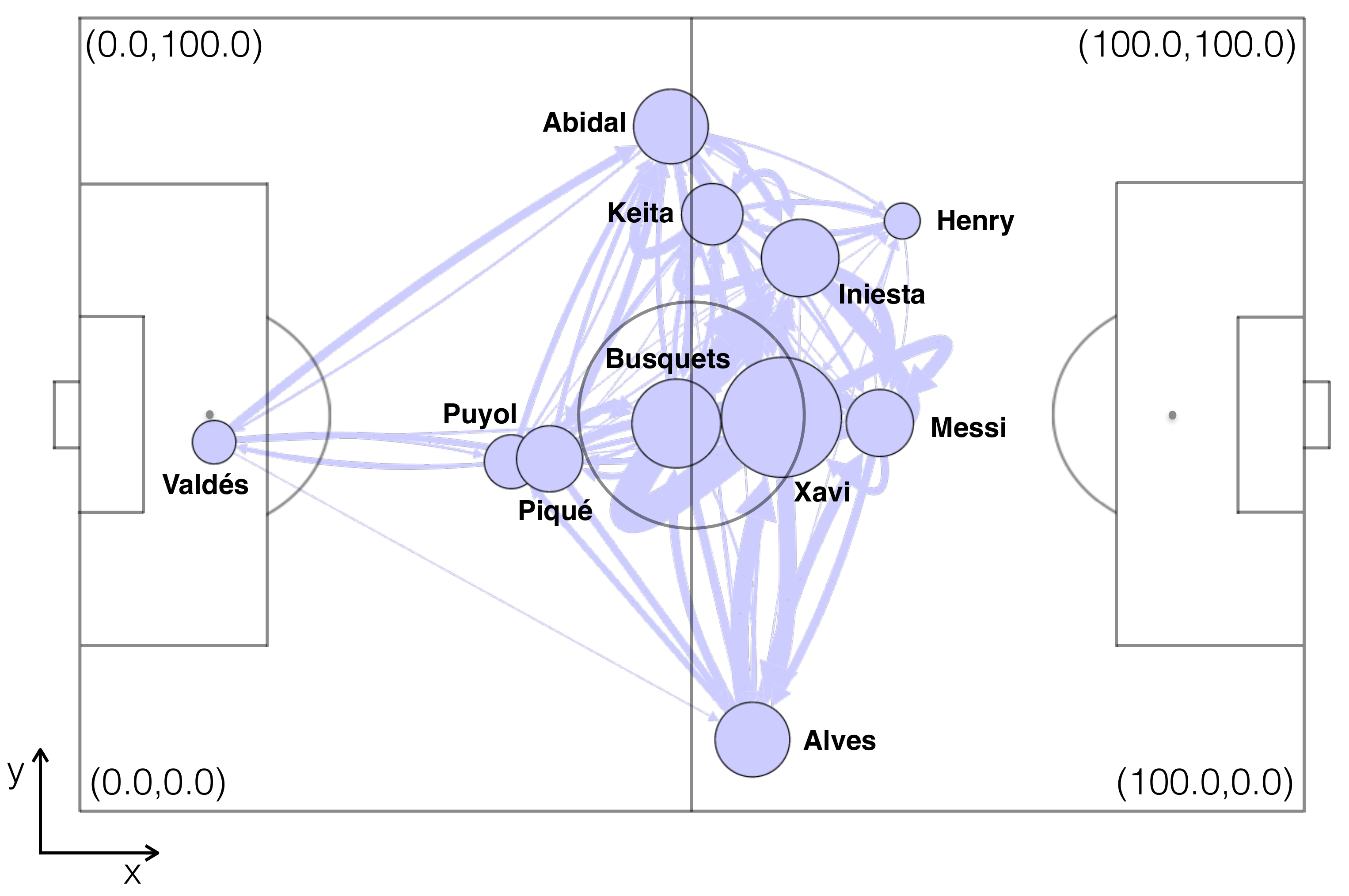}
 \caption{
Schematic illustration of a football passing network. In the plot, players are represented by circular nodes, whose size is proportional to their eigenvector centrality, a mesure of importance in the
network structure. The position of each player is given by the average of the positions of all passes made by the player along the match. The width of the links is proportional to their weights, which account 
for the number of passes between players. Note that links are unidirectional. In this example, we plot the average passing network of the match between F.C. Barcelona and 
Real Madrid, played during the season 2009/2010 at Santiago Bernabeu Stadium. Datasets leading to the passing network were provided by Opta.
}\label{fig:f01}
\end{figure}

\section*{Results}

\subsection*{Average Passing Networks}

Figure \ref{fig:f01} shows an example of a football passing network, in this case the average network of FCB against Real Madrid in the season 2009/2010. 
Note that links are unidirectional (from player A to player B) and weighted according to the number of passes between players. In the figure, nodes (i.e., players) are
placed in the average position from where their passes were made and the width of the links is proportional to the number of passes between players. Also note that both the $x$ and $y$ coordinates of the field are bounded
between [0,100] and are measured in ``field units" (f.u.), since not all fields have exactly the same dimensions. Finally, the radius of the nodes is proportional to their importance in the passing network, quantified
by means of the eigenvector centrality (see Methods).

First, we analyzed the {\it average passing networks} of all matches played by FCB during season 2009/2010 ($38$ in total), obtaining the networks of FCB and their rivals. Specifically, we obtain 2 average
passing networks for each match (1 per team), both of them including all passes and positions along the match and projecting them into a single network for each team. See the Methods section for 
details about the construction of average passing networks.
Previous literature about average passing networks has shown that they reveal information about the way a team is organized \cite{lopez2012} and are
also related with team performance \cite{cintia2015}. 

\begin{figure}[!b]
 \centering
 \includegraphics[width=0.9\textwidth]{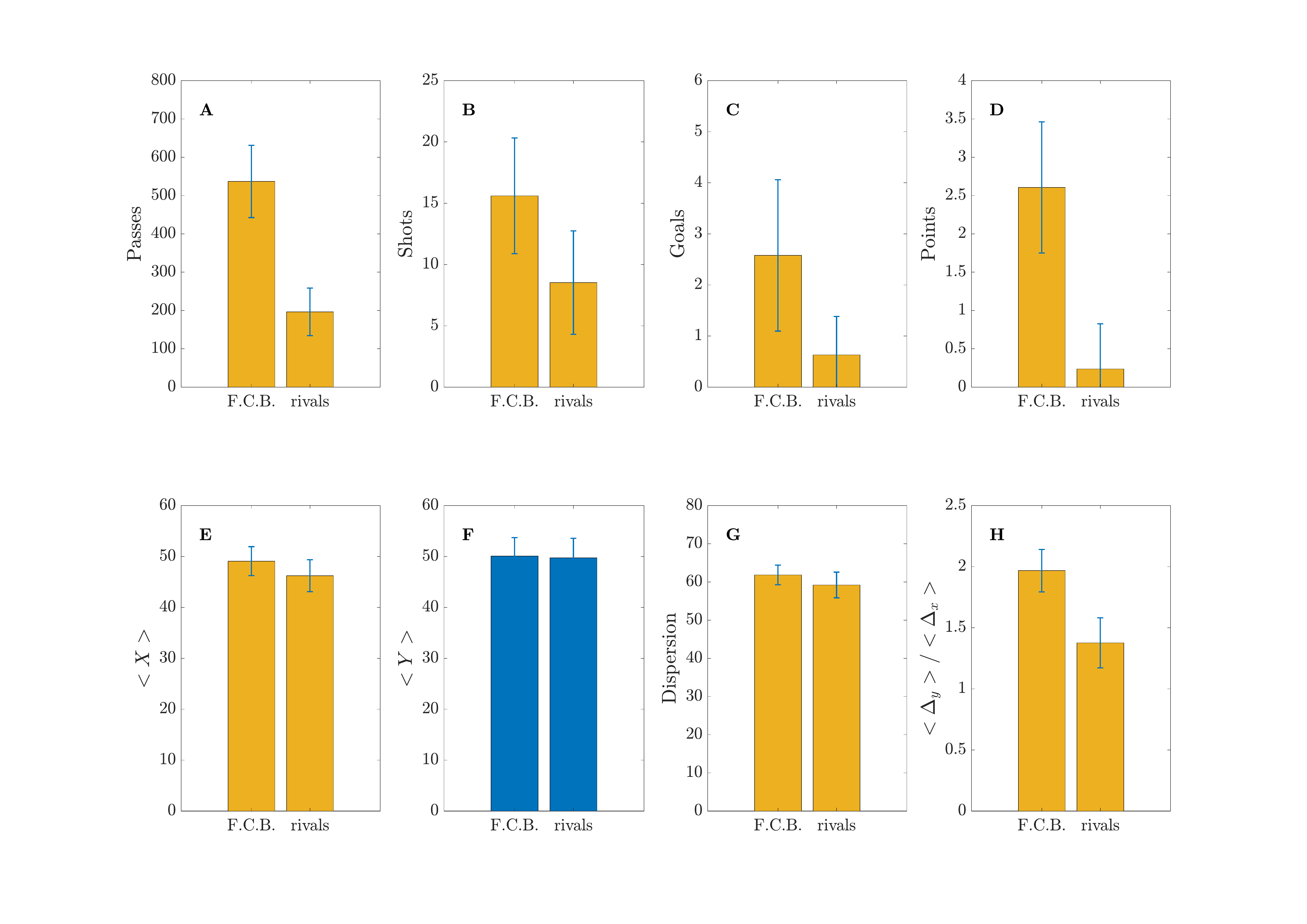}
 \caption{
Comparison of 8 classical football metrics. In all plots, left bars are the average (during the whole season) of a given metric for FCB, while right bars correspond to the average of the rivals in the matches
played against FCB. Metrics are, specifically: (A) number of passes, (B) number of shots, (C) number of goals, (D) number of points at the end of the season, (E) $x$-coordinate of the network centroid $\langle X \rangle$, (F) $y$-coordinate of the network centroid $\langle Y \rangle$, (G) the spatial dispersion (in field units) of the players around the network centroid and (H) the advance ratio $\langle \Delta_y \rangle / \langle \Delta_x \rangle$, obtained as the ratio between the total length $\langle \Delta_y \rangle$  of the $y$-coordinate of all passes divided by the total length $\langle \Delta_x \rangle$ of the $x$-coordinate, both distances in field units. Direction $x$ is towards the goal, while direction $y$ is parallel to the opponents goal (see axis of Fig. \ref{fig:f01}). Parameters having statistically significant differences between FCB and its rivals are plotted in yellow.
}\label{fig:f02}
\end{figure}

Figure \ref{fig:f02} shows the comparison between 8 different parameters obtained for FCB and its rivals. 
Four of them, (a) the number of passes $L$, (b) the number of shots to goal $M_{shots}$, (c) the number of goals $M_{goals}$ and (d) the number of points $M_{points}$
(at the end of the season) are classical metrics of the team performance. Note that, in order to compute these 4 variables, there is no need to obtain and analyze the network
structure of each team, despite some of them (i.e., the number of passes) can affect the organization of the passing networks. The other 4 parameters of Fig. \ref{fig:f02} are related to the spatial properties of the networks:
(e) $x$-coordinate of the network centroid $\langle X \rangle$, (f) $y$-coordinate of the network centroid $\langle Y \rangle$, (g) dispersion of the position of the players around the 
network centroid $NC_{disp}$ and (h) average ratio between the passing distance parallel and perpendicular to the opponent's goal $\langle \Delta_y \rangle / \langle \Delta_x \rangle$ 
(see Methods for details). Left bars in all plots correspond
to the average values of these metrics for all matches of FCB along the season, 
while right bars are the same metrics obtained for the rivals at the same matches. FCB is always averaged with itself, while all other teams are averaged together, the reason being that we are only interested in observing differences between the FCB and all other teams. Error bars account 
for the standard deviations of each metric. Plots in yellow highlight statistically significant differences (see Methods for details about the statistical analysis).

As we can see in Fig. \ref{fig:f02}A, the number of passes made by FCB is much higher than the average of their rivals. This fact is a consequence of Guardiola's playing style, focused on keeping the ball as much as possible ({\it ``In football, I am very selfish: I want the ball for me", ``take the ball, pass the ball" \cite{duncan}}). The high number of passes unavoidably leads to passing networks with links that have higher weights and, 
as we will see, this fact will have consequences on the network parameters. The number of shots to goal is also higher in FCB (Fig. \ref{fig:f02}B), leading to a higher 
number of goals (Fig. \ref{fig:f02}C) and, ultimately to a high number of points accumulated during the analyzed matches (Fig. \ref{fig:f02}D). 
In fact, FCB won the league with $99$ points ($31$ wins, $6$ ties and only $1$ loss). Note that these four metrics (passes, shots, goals and points), specially the last three, are traditionally considered as indicators of the team performance, thus revealing that FCB was the best team during season 2009/2010.

Bottom plots of Fig. \ref{fig:f02} are related with the spatial features of Guardiola's team. The $\langle X \rangle$ and $\langle Y \rangle$ average coordinates of all passes made during the match define the {\it network centroid} 
(or the network {\it center of mass}). We can observe in Fig. \ref{fig:f02}E how FCB played closer to the opponents goal ($\langle X\rangle _{FCB} > \langle X\rangle _{rivals}$), while no differences
are found at the $\langle Y\rangle $ coordinate (Fig. \ref{fig:f02}F), indicating no preference for any of the sides of the pitch.
Interestingly, the dispersion of the position of the players around the centroid (see Methods) is slightly higher for FCB, which indicates that the area covered by the initial position of the passes made by all players
is wider (Fig. \ref{fig:f02}G). Finally, it is worth analyzing the {\it ratio of advance} $\langle \Delta_y \rangle / \langle \Delta_x \rangle$, which is an 
indicator of the direction of the passes of a team, since the $\Delta_y=y_2-y_1$ of a pass is the difference
between the $y$-coordinates at the final ($y_2$) and initial points ($y_1$) of a pass, while $\Delta_x$ is defined, accordingly, for the $x$-coordinate. In Fig. \ref{fig:f02}H, we can 
observe how FCB has a ratio of advance much higher than the rivals, which reveals that 
passes are more parallel to the opponent's goal than the rest of the teams. Note that this metric is independent from the number of passes, and it is an indicator of how ``direct" the game of a team
is. Clearly, FCB is not concerned about advancing directly towards the goal, but on moving the ball in parallel, probably to find the most adequate moment to advance.

\begin{figure}[!b]
 \centering
 \includegraphics[width=0.8\textwidth]{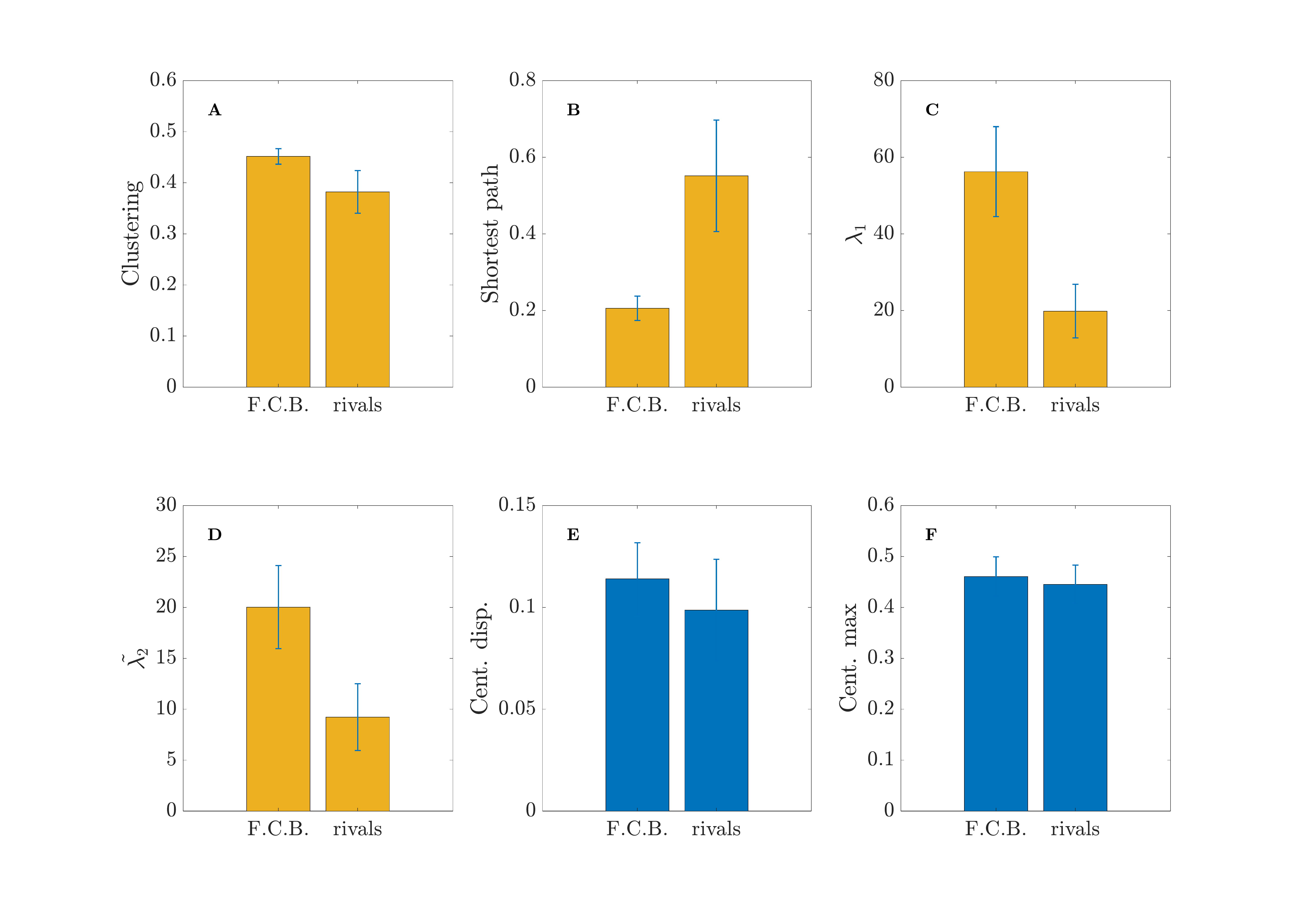}
 \caption{
Comparison of 6 network parameters. In all plots, left bars are the average (during the whole season) of a given parameter for FCB, while right bars correspond to the average of the rivals in the matches
played against FCB. Parameters are, specifically: (A) clustering coefficient $C$, (B) shortest-path length $d$, (C) largest eigenvalue $\lambda_{1}$ of the connectivity matrix $A$, 
(D) algebraic connectivity $\tilde{\lambda}_2$ of the Laplacian matrix $\tilde{L}$, (E) dispersion of the players' centrality and (F) maximum player centrality. See Methods section for details about the explanation (and calculation) of all network parameters. Parameters having statistically significant differences between FCB and their rivals are plotted in yellow.
}\label{fig:f03}
\end{figure}

But, how is the structure of the average passing networks? And, more importantly, are there differences between FCB and the rest of the teams?
Figure \ref{fig:f03} shows the comparison of $6$ parameters directly related with the topological organization of the average passing networks (see Methods for a detailed description of all these
network parameters). In Fig. \ref{fig:f03}A, we plot the {\it clustering coefficient} $C$, which is related to the amount of triangles created between any triplet of players. Clustering coefficient is an indicator of the local robustness of networks \cite{newman2010}, since when a triangle connecting three nodes (i.e. players) exists, and a link (i.e., pass) between two nodes is lost (i.e., not possible to make the pass), 
there is an alternative way of reaching the other node passing through the other two edges of the triangle. In football, the clustering coefficient mesures the triangulation between three players. As we can observe
in Fig.\ref{fig:f03}A the value of $C$ is much higher in FCB, which reveals that connections between three players are more abundant than at its rivals.
The {\it average shortest path} $d$ is an indicator about how well connected are players inside a team. It measures the 
``topological distance" that the ball must go through to connect any two players of the team. Since the
links of the passing networks are weighted with the number of passes, the topological distance of a given link is defined as the inverse of the 
number of passes. The higher the number of passes between two players,
the closer (i.e., lower) the topological distance between them is. Furthermore, since it is the ball that travels from one player to any other, it is possible to find the shortest path between any pair of players by computing
the shortest topological distance between them, no matter if it is a direct connection or if it involves passing through other players of the team. Finally, the 
average shortest path $d$ of a team is just the average of the shortest
path between all pairs of players. As we can observe in  Fig.\ref{fig:f03}B, the shortest path of FCB is much lower than their 
rivals, which reveals that players are better connected between them. As we will discuss later, note that this fact
could be produced by the network organization or just being a consequence of having a higher number of passes, which reduces 
the overall topological distance of the links and, consequently, the value of $d$. 

Figure \ref{fig:f03}C shows the comparison between the {\it largest eigenvalue} $\lambda_{1}$ of the connectivity matrix $A$ (also known as the weighted adjacency matrix), whose elements $a_{ij}$ contain the number of 
passes between players $i$ and $j$ \cite{newman2010}. The largest eigenvalue has been used as a quantifier of the network 
strength \cite{aguirre2013}, since it increases with the number of nodes and links (see Methods). 
As expected (due to the high number of passes), the largest eigenvalue $\lambda_{1}$ of FCB is much higher than the corresponding values of their rivals. This metric reveals the higher
robustness of the passing network of Guardiola's team, which indicates that an eventual loss of passes would have less consequences in F.C. Barcelona than in the rest of the teams.

It is also worth analyzing the behavior of the second smallest eigenvalue $\tilde{\lambda}_2$ of the Laplacian matrix $\tilde{L}$, also known as the {\it algebraic connectivity} (see Methods). 
The value of $\tilde{\lambda}_2$ is related to several network properties. 
In synchronization, networks with higher $\tilde{\lambda}_2$ require less time to synchronize \cite{almendral2007} and in diffusion processes, the time to reach equilibrium also goes with the inverse of $\tilde{\lambda}_2$.
In the context of football passing networks, $\tilde{\lambda}_2$ can be interpreted as a metric for quantifying the division of a team. The reason is that low values of $\tilde{\lambda}_2$ indicate that a network
is close to be split into two groups, eventually breaking for $\tilde{\lambda}_2=0$. In this way, the higher the value of $\tilde{\lambda}_2$ the more interconnected the team is, being a measure of structural cohesion.
In Fig. \ref{fig:f03}D, we have plot the comparison of $\tilde{\lambda}_2$, which reveals that FCB attacking and defensive lines are more intermingled, leading to a $\tilde{\lambda}_2$ higher than their rivals.

Finally, Figs. \ref{fig:f03}E-F show how {\it centrality} (i.e., the importance of the players inside the passing network) is distributed along the team, a metric calculated by means of the eigenvector related
to the largest eigenvalue of the connectivity matrix (see Methods). Figure \ref{fig:f03}E contains the average dispersion of centrality and Fig. \ref{fig:f03}F shows the highest value of a single player.
In both cases, differences are not statistically significant to support evidences of a different centrality distribution between FCB and the rest of the teams.

\subsection*{Temporal evolution of the network metrics}

As we have seen in the previous Section, average passing networks show differences between the organization of FCB and its rivals. However, these difference may be interpreted as a consequence
of the higher number of passes between Barcelona players, which could lead to statistically significant differences in a diversity of network metrics, namely, a 
reduction of the average shortest path $d$ and an increase of the clustering coefficient $C$, largest eigenvalue $\lambda_1$ and algebraic connectivity $\tilde{\lambda}_2$. 

In view of these results, two questions must be addressed before any interpretation: (i) Is just the number of passes behind the differences of the network parameters? and (ii) is it enough to look at the average values of the network metrics?
To address both issues, we have conducted a complementary study where passing networks are constructed in a different way. On the one hand, we are going to define passing networks as non-static entities,
 thus evolving in time, and we will track the evolution of their parameters. On the other hand, we are going to exclude the importance of the number of passes, in order to just focus on the topological organization
 of the networks.
 With these two objectives in mind, we construct the {\it $l-$pass networks} of a team as the networks containing $l$ consecutive passes, with $l<<L$, being $L$ the total number of passes during the match.
 In our study, we set $l=50$, since it is a value low enough to allow a tracking of the network evolution along the match and, at the same time, high enough to guarantee the creation of a network between players
 (too low values of $l$ would lead to networks with disconnected components). Therefore, we obtain the {\it $50$-pass networks} in the following way: (i) we construct the network of the first
 $50$ passes of a team since the beginning of the match, (ii) we calculate its parameters, (iii) we dismiss the oldest pass and include (sequentially) a new one, (iv) we recalculate the network parameters  and (v) we
 repeat the procedure until the last pass of the match is included.
 
\begin{figure}[!b]
 \centering
 \includegraphics[width=0.8\textwidth]{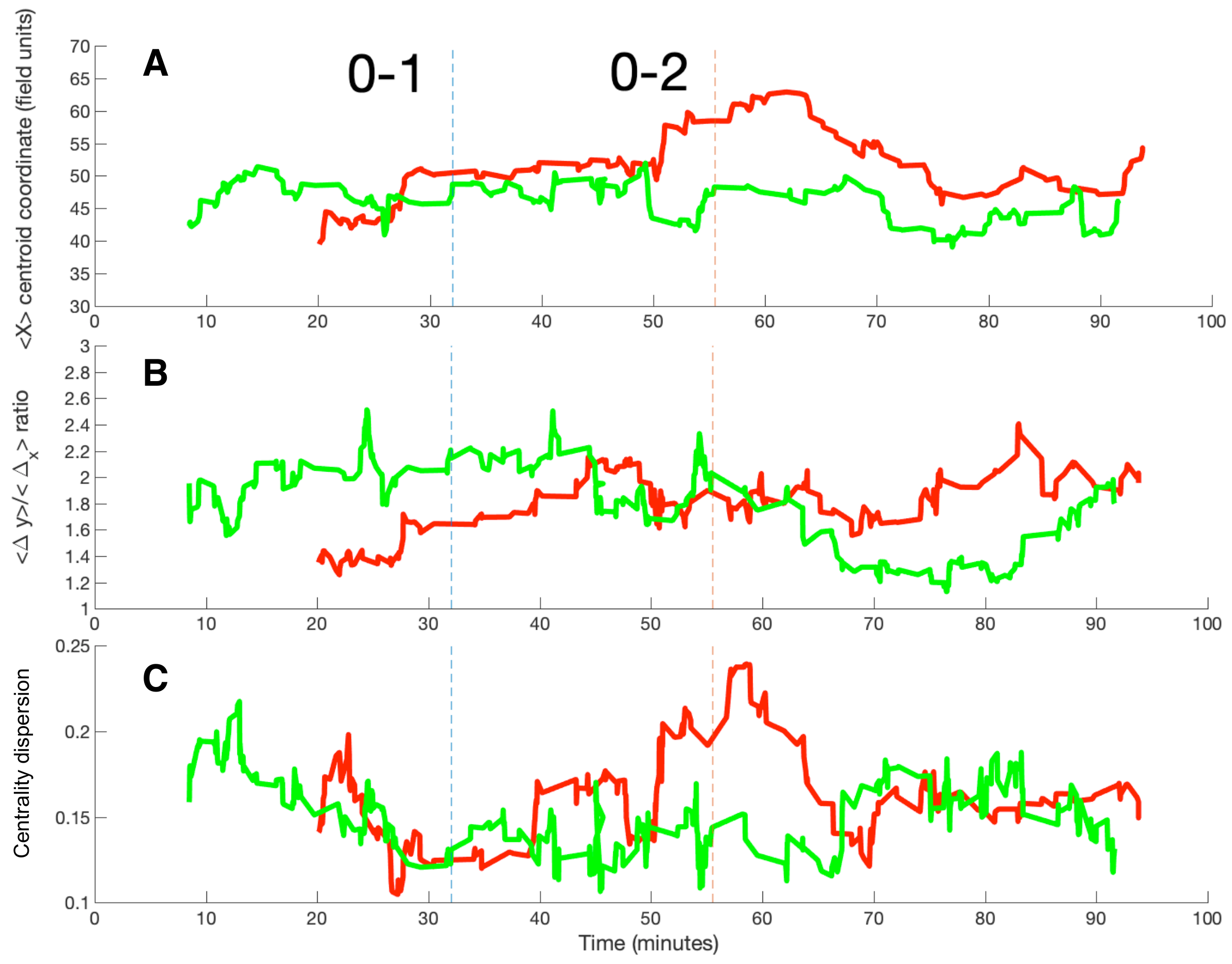}
 \caption{
Real Madrid (red lines) vs. F.C. Barcelona (green lines), season 2009/2010 (final result: $0-2$).Temporal evolution of the network parameters: (A) $\langle X \rangle$ coordinate of the networks' centroid, (B) ratio of advance $\langle \Delta_y \rangle / \langle \Delta_x \rangle$ and (C) the centrality dispersion $EC_{disp}$. Vertical dashed lines indicate the two moments
 at which FCB scored a goal (Real Madrid did not score).} \label{fig:f04}
\end{figure}
 
Note that $50$-pass networks contain exactly the same number of passes for both teams and, thus, any difference between network metrics can not be attributed to the total number of passes. In addition,
also note that metrics evolve in time and their values can be related to a certain moment of the match. However, it is also important to remark that the time required to construct
a $50$-pass network can differ from team to team.

Figure \ref{fig:f04} shows an example of the evolution of $3$ parameters of the $50$-pass networks of two teams along a match, specifically, the $\langle X \rangle$ coordinate of the centroid (A), the ratio of advance $\langle \Delta_y \rangle / \langle \Delta_x \rangle$ (B), and the dispersion of the network centrality (C). Parameters are calculated, for both teams, during the match between Real Madrid (red lines) and FCB (green lines), whose final score was $0$-$2$. Vertical lines indicate the moment
at which a goal was scored. Figure \ref{fig:f04}A shows how the position of the team moves forward and backward during the match. In this particular
case, Real Madrid plays, most of the time, more advanced than FCB, which did not lead to an advantage in the result. Note how the centroid of FCB seems to be more stable, while Real Madrid has higher fluctuations,
arriving to its maximum value around minute $63$. Also note how FCB is the first team to construct the $50$-pass network around minute $9$, while Real Madrid required $20$ minutes.

In Fig. \ref{fig:f04}B, we plot the ratio of advance of the 
$50$-pass networks of both teams. Again we can see fluctuations of the parameter during the match. Specifically, FCB has a highest value during the first part of the match. However, we can observe how Real Madrid increases its advance ratio as time goes by, eventually overcoming FCB during the second half.

\begin{figure}[!b]
 \centering
 \includegraphics[width=1.0\textwidth]{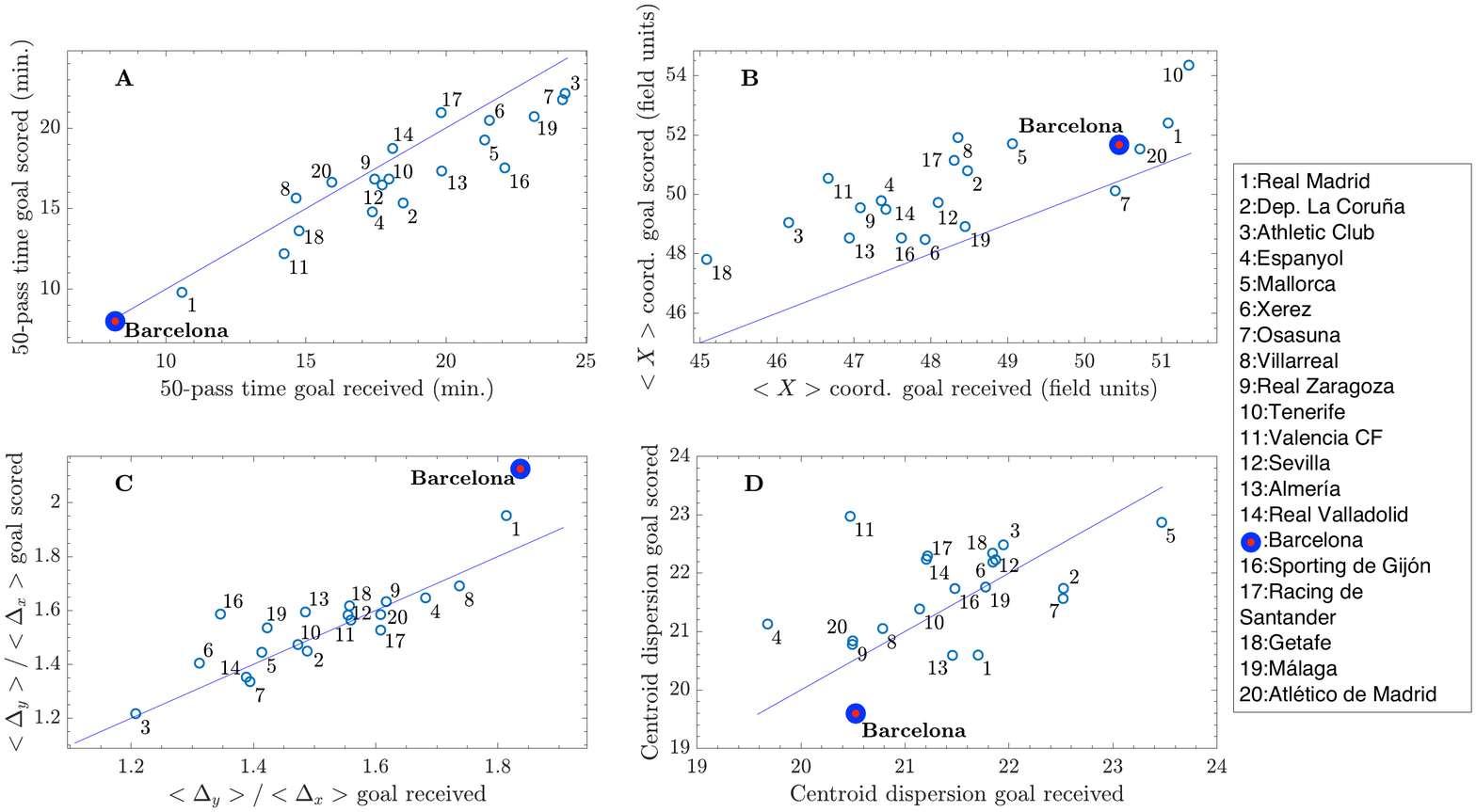}
 \caption{
Temporal and spatial metrics change before scoring/receiving a goal: (A) time required to construct a 50-pass network $t_{net}$, (B) position of the X coordinate of the 50-pass network centroid, (C) $\langle \Delta_y \rangle / \langle \Delta_x \rangle$ advance ratio and (D)
dispersion of the distance of the players with regard to the centroid. Metrics are obtained for all teams and are shown in a two-dimensional plot, where the horizontal axis corresponds to the value of a metric
when the team receives a goal and the vertical axis is the same metric obtained when the team scores a goal. Solid lines correspond to the function $y=x$, helping to identify wether a given parameter increases
or decreases when a goal is scored/received. Each point represents the average along the whole season.
}\label{fig:f05}
\end{figure}

Finally, Fig. \ref{fig:f04}C shows the fluctuations of the centrality dispersion of the players of both teams. We can observe how Real Madrid has a strong increase of the centrality dispersion between minutes
50 and 70, which seems to be related with the period where the centroid of the team advances towards FCB's goal (see Fig. \ref{fig:f04}A). This change of the centrality distribution could
be related to a change of the style of playing. Since centrality dispersion increases, there is a higher heterogeneity in the importance of the players in the passing networks, 
which could be related to the fact that a few players are taking the lead of the team. 
However, this change in the organization of the passing network does not seem to be effective, since the second goal of FCB comes around to the maximum of
centrality dispersion.

The fact that network metrics change during the match increases the complexity of the study. It is expected that several factors may influence the fluctuations of the network parameters (a goal, a substitution, physical condition, etc... ) and, furthermore, not all teams may behave in the same way. 
From the diversity of factors, here we are going to focus on the particular
organization of each team before a goal. 
With this aim, we have analyzed the value of the network parameters, for all teams, before scoring/receiving a goal. Our purpose is to detect the 
existence of differences in the network metrics and identify
those parameters that change before scoring or receiving a goal.

Figure \ref{fig:f05} shows the average values of $4$ temporal and spatial metrics obtained before scoring/receiving a goal (during season 2009/2010). 
The diagonal line ($y=x$) helps to identify those metrics that behave differently when scoring
or receiving a goal. In Fig. \ref{fig:f05}A we can observe how FCB is the team requiring less time to construct the $50$-pass network, both 
when scoring or receiving a goal. In fact,
as indicated by the diagonal line, it takes approximately the same time in both cases. On the opposite side, we find Athletic Club and
Osasuna, both teams characterized by a direct game towards the opponents's goal.
Concerning the  $\langle X \rangle$  position of the centroid, we can observe in Fig. \ref{fig:f05}B that, despite having a high value, FCB is not the team that constructs its network closest to the opponent's goal, since it is overcome
by Real Madrid and Tenerife. Note that Tenerife ended up the season in the last position, which indicates that playing forward it is not a sufficient condition to achieve good results.
However, it is also worth noting that all teams, with the only exception of Osasuna, are placed above the line given by the function $\langle X \rangle_{scored}=\langle X \rangle_{received}$. This fact reveals that when a team scores a goal is, in average, playing more advanced than
when it receives it. In Fig. \ref{fig:f05}C we have compared the ratio of advance $\langle \Delta_y \rangle / \langle \Delta_x \rangle$ of all teams, showing that Barcelona is not only the team with the highest value (both when scoring and receiving a goal) but also the one deviated the most from the the diagonal line. In this way, FCB is the team that increases the most its probability of scoring a goal when increasing the ratio of advance. Finally, Fig. \ref{fig:f05}D shows
the average dispersion of the position of the players around the centroid coordinates of the $50$-pass network. Interestingly, we can observe how FCB is one of the teams with lower dispersion 
of {\it La Liga} and, furthermore, the dispersion increases before a goal is received, indicating that FCB performs better when players are closer to the network centroid.

\begin{figure}[!t]
 \centering
 \includegraphics[width=1.0\textwidth]{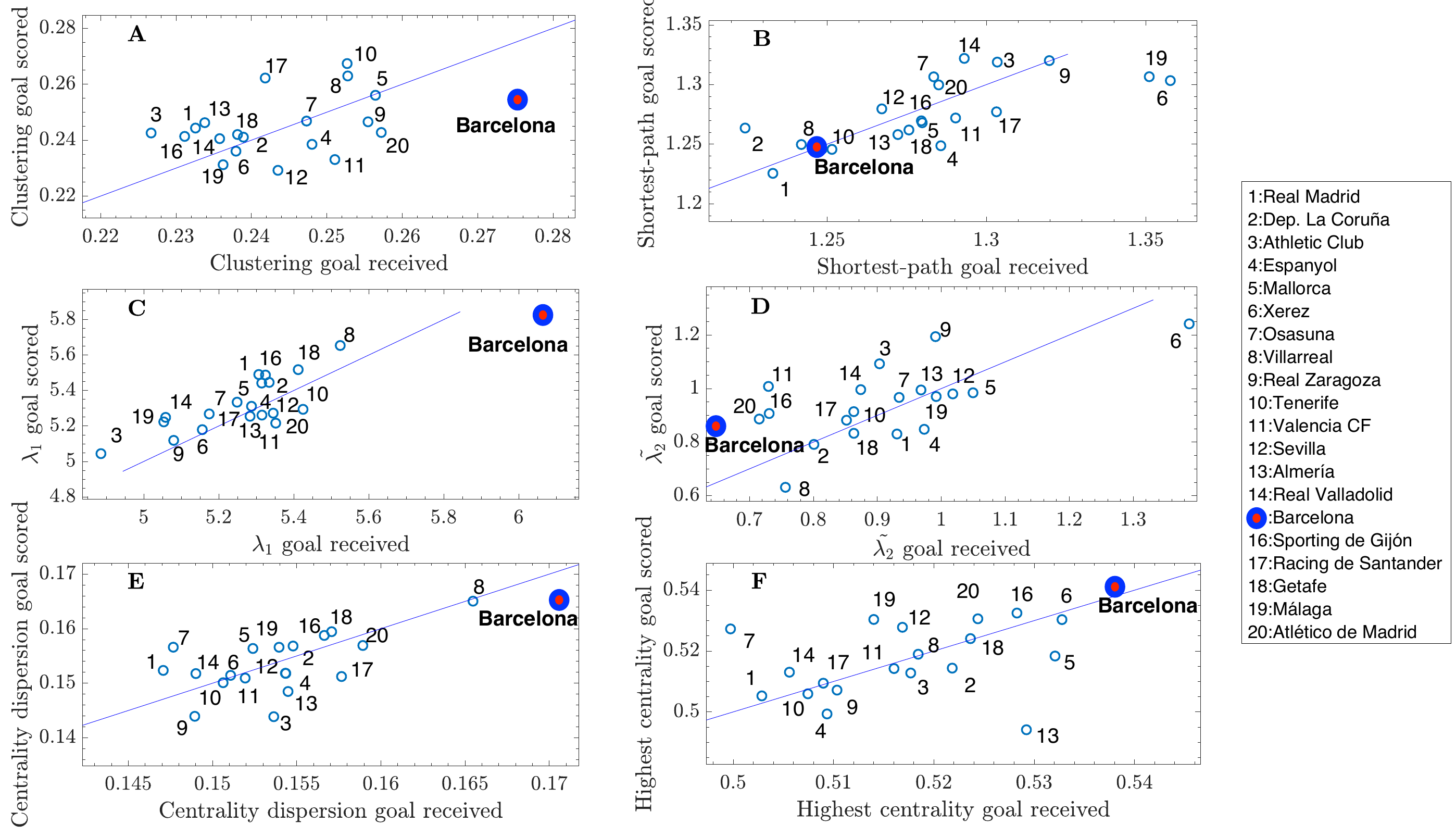}
 \caption{
Network parameters depend on scoring/receiving a goal. (A) clustering coefficient $C$, (B) average shortest-path $d$, (C) largest eigenvalue $\lambda_1$ of the connectivity matrix, (D)
algebraic connectivity $\tilde{\lambda}_{2}$, (E) centrality dispersion $EC_{disp}$ and (f) highest eigenvector centrality $EC_{max}$. Parameters are obtained for 
all teams and are shown in a two-dimensional plot, where the horizontal axis corresponds to the value of a metric
when the team receives a goal and the vertical axis is the same metric obtained when the team scores a goal. Solid lines correspond to the function $y=x$, helping to identify wether a given parameter increases
or decreases when a goal is scored/received. Each point represents the average along the whole season.
}\label{fig:f06}
\end{figure}

Figure \ref{fig:f06} shows, in a similar way, the values of $6$ different network parameters obtained for all teams (during the whole season). Interestingly, FCB has the highest values of the 
league at 4 of them: The clustering coefficient (Fig. \ref{fig:f06}A), the largest eigenvalue of the connectivity matrix (Fig. \ref{fig:f06}C), the centrality dispersion (Fig. \ref{fig:f06}E)  and the 
highest centrality of a player (Fig. \ref{fig:f06}F). High values of these four metrics are related to strong and robust networks: (i) a high 
clustering coefficient is an indicator of local robustness \cite{watts1998,newman2010}, (ii) the largest eigenvalue $\lambda_1$ is also 
an indicator of global robustness \cite{aguirre2013}; when the number of nodes and links are the same, $\lambda_1$ increases 
when important players are, in turn, connected between them, (iii) a high centrality dispersion together with a high value of maximum centrality are indicators of heterogeneity 
in the network structure, and heterogeneous networks are know to have strong resilience against random failures \cite{albert2000} (i.e., the loss of weight of the links, due to lost passes, 
would have less impact on the overall structure).

At the same time, the analysis shows low values at other $2$ metrics: the shortest-path length $d$ (Fig. \ref{fig:f06}B) and the algebraic connectivity $\tilde{\lambda}_{2}$ (Fig. \ref{fig:f06}D). 
In this case, having a low shortest-path length
is an indicator of a better connection between players, since the ball can travel from a player to any other in a lower number of steps. Finally,
 it is interesting to note that FCB has one of the lowest algebraic connectivities,
which is an indicator of structural integration. Low values of $\tilde{\lambda}_2$ reflect that the team is more split into two different groups.  
Note that, when the algebraic connectivity $\tilde{\lambda}_{2}$ is calculated from the average connectivity matrix 
 (Fig. \ref{fig:f03}D), FCB has a value higher than their rivals, reflecting a higher cohesion of the whole team. However, 
 when it is computed from the $50$-pass networks, FCB algebraic connectivity is one of the lowest. 
 A possible explanation is that cohesion of the team may be grounded on a higher number of passes between players, and not on the topological organization of the network.

\section*{Discussion}

\subsection*{What passing networks tell us about FCB}
As we have seen, using Network Science to analyze football passing networks gives a new perspective that allows distinguishing between different teams and relating
network properties to the teams particular style of playing. Here, we have made use of these metrics to characterize the passing networks of Guardiola's Barcelona, focusing
on the season 2009/2010 of the Spanish national league, one of the years where FCB was considered to reach its top in terms of playing style and trophies.
When passing networks are constructed as a simple addition of all passes made between players during the match, statistically significant differences between the passing networks of
FCB and its rivals arise. Specifically, the clustering coefficient, the shortest-path, the largest eigenvalue of the connectivity matrix and the algebraic connectivity, always have {\it ``better"} values in the Catalan team.
The term ``better" refers to the fact that differences in these network properties are related with a higher local resilience against the loss of passes (due to a higher clustering), a lower number of steps to connect
any two players of the teams (due to a lower shortest-path length) and a higher connectedness between the whole team, as indicated by a higher largest eigenvalue of the connectivity matrix and a higher algebraic 
connectivity. 

However, it is worth looking beyond the differences in the network metrics and trying to find the reasons behind them. When focusing on the number of passes made by FCB we can, first, observe that it is much higher than their rivals and, second, that the advance ratio, measuring the percentage of distance that the ball advances parallel to the opponent's goal is also much higher.
Concerning the latter, note that the advance ratio is not related to the number of passes and, therefore, there is not an obvious reason why it should influence network parameters. However, the number of passes has,
 indeed, crucial consequences on any quantitative analysis using Network Science. The fact that we are comparing networks with the same number of nodes (eleven) but links with different weights (number of passes) 
 has unavoidable consequences on the network parameters. For example, since the ``topological" distance between two directly connected players is given by the inverse of the number of passes between them, the
 higher the average number of passes of a team, the lower topological distance between their players. Despite being obvious, a reasonable conclusion of the study is that increasing the number of passes benefits the general properties of passing networks. However, comparing the properties of two networks with different number of passes hinders the role played by the network topology itself, i.e., we can not say that a network is better organized, since we can not separate the effect of the number of passes (``quantity") from that of the topology of the network (``quality").
 
A second issue related to the number of passes is possession. Note that the number of passes is intimately related to the possession a team has. 
A team with higher possession will unavoidably have more passes and that is exactly what
FCB, under the guidance of Guardiola, is doing. But to what extend can we relate possession to the particular organization of FCB passing networks? Are the reported values of its network
parameters just a consequence of having the ball more time?
 
A third issue arises when trying to interpret the results of the averaged passing networks. Since, as we have seen, network organization and, consequently, network parameters, 
are continuously evolving during the match, considering the sum of all passes may hide interesting information about how different crucial events influence a team's style of 
 playing, such as a scored/received goal, a substitution or, simply, the fatigue of the players as time goes by.
 
In order to overcome these three issues and, particularly, to exclude the influence of the number of passes (or possession) and, at the same time, accounting for the evolution of the network topology we studied the properties of the $50$-pass networks.
 A part from the benefits of tracking their temporal evolution, $50$-pass networks contain exactly the same number of nodes and links for the two teams playing a match, which allows a direct comparison of the network organization,
 no matter what the final number of passes of each team is. However, the tracking of the parameters of the $50$-pass network shows that network parameters are in continuous evolution, which increases the complexity
 of the analysis. Here, we focused on the state of the passing network just before scoring/receiving a goal, which allows to extract information about what are the network properties associated to the ability
 of a team to score/receive a goal. With such an approach we were able to complement the information extracted from the averaged passing networks, obtaining a more detailed profile of Guardiola's team.
 These results reinforced all the conclusions drawn by analyzing average passing networks and included the following additional information about FCB: 
 \begin{enumerate}
 \item It is the team that requires the shortest time to construct $50$-pass networks, and this time remains unaltered when scoring/receiving a goal, 
 \item It is the team with the highest advance ratio (i.e., the team that plays the most horizontal
 to the opponent's goal) and this metric is specially high before scoring a goal, 
 \item The dispersion of the players around the network centroid is the lowest but significantly increases before receiving a goal, 
 \item The clustering coefficient is higher when receiving goal than when a goal is scored, 
  \item The shortest-path is one of the lowest and does not depend on scoring/receiving a goal, 
 \item The largest eigenvalue of the adjacency matrix, measuring
 the strength of the network is the largest, and significantly increases before receiving a goal, 
 \item The algebraic connectivity, measuring the cohesion between groups of players,  decreases
 before receiving a goal (i.e., the interplay between groups is reduced), 
 \item The highest centrality acquired by a single player and the centrality dispersion are the highest, which indicates that the importance of players in the FCB network is not evenly distributed, with one player, Xavi,
 being the hub of the passing networks. 
 \end{enumerate}
 
Note that all these patterns refer exclusively to FCB, while passing networks corresponding to other teams behave
  in its own way. Therefore, one of the conclusions we can draw from
 Fig. \ref{fig:f05} and Fig. \ref{fig:f06} is that variables of each team before scoring/receiving a goal behave in a particular way.  
 For example, as we 
 can see in Fig. \ref{fig:f05}D, the dispersion of FCB's players around the position of the network's centroid is higher when a goal is received, indicating that when players
 are more separated from the centroid, the risk of receiving a goal increases. However, if we look at the same parameter for Valencia CF, we can observe that the behaviour is just the
 opposite, and higher dispersions around the team's centroid (note that in this case players are occupying more field) are reported when scoring a goal. 

Finally, it is worth mentioning the limitations and risks of our study. As we have seen, computing the parameters related to the average passing networks
 gives interesting, but limited, information about the way a team is organized. As shown in Fig. \ref{fig:f04}, there exist strong fluctuations on the network parameters during a match
 and defining 50-pass networks is a reasonable option to capture the evolution of the structure of passing networks. However, there are associated issues and alternatives that may be explored
 in further studies. For example, the length of the 50-pass networks could be adapted to capture the ``momentum" of the match, which may change from team to team or just due to the events 
 occurring during the match. Therefore, it would be interesting to find a way of defining the most adequate time windows and how the length of these windows are related to the particular style of
 playing a team has. Another limitation is related to the causes of the parameter fluctuations, since they can have different origins (a goal, a substitution, fatigue, etc...). It would be extremely
 useful to identify all possible variables affecting the network organization and compute the network parameters after these particular events occur, trying to identify what are those variables
 that crucially change the style of playing of a given team.

\subsection*{An interpretation within a football framework }

Going beyond passing networks, the strategy of having the ball most of the time leads, in general, to controlling the game by creating a dynamical context 
to which the opposing team needs to adapt and, in particular, gave FCB a systematic superiority that led to an increase of the scoring opportunities. 

Firstly, FCB defense can lengthen or shorten the space by 
moving the line of defenders forwards or backwards. In other words, it can play with the occupied length of the field and use the {\it off-side area} in its favor. 
The fact that FCB network centroid was advanced (in average) compared to its rivals (see Fig. \ref{fig:f02}E), left the opponents a smaller area of the pitch and fewer playing options. 

Furthermore, Barcelona organized the team into {\it ``situational areas"} around the ball, which comprised the commitment of five or six players. Inside these areas, 
the team must overcome a challenge, i.e., either play the ball or recover it, leading to a division of the game into two phases. Players organized spontaneously inside 
these situational areas (and, as we will discuss, after training these situations), communicating with each other and exchanging physical, verbal, and 
motor-related signals. This way of modulating the playing field into building blocks leads to more playing patterns, resulting in more 
different options to overcome the rival. 

\begin{figure}[!h]
 \centering
 \includegraphics[width=0.7\textwidth]{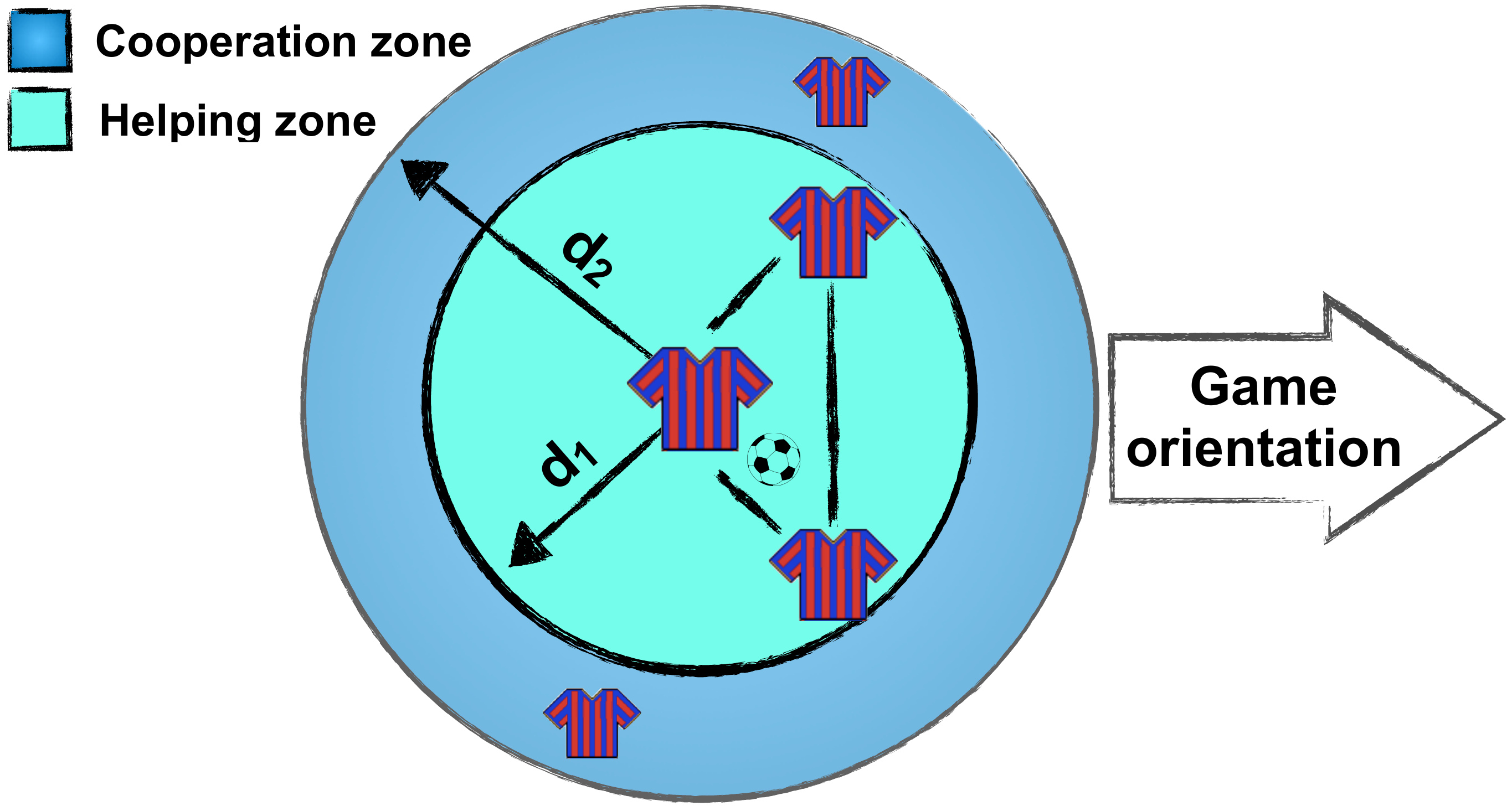}
 \caption{ F.C. Barcelona organization during {\it ``attacking phases"}. Four players organize around a fifth player who is having the ball. Two concentric circles around the ball define the helping
 zone (radius $d_1$) and the cooperation zone (radius $d_2$). Passes between players in the helping zone are promoted. The defensive phase is organized in a similar way, but in this case, pressing
 the opponent who has the ball.
}\label{fig:f07}
\end{figure}

Specifically, during the {\it attacking phase} (see Fig. \ref{fig:f07}), the player with the ball had a helping area \#$1$ (also known as the helping zone) with two players 
forming possible triangles within a distance $d_1$ of 10 to 15 meters. At the same time, there was a co-operation area (\#$2$) with two more players 
(one slightly forward and the other covering the back) occupying a wider radius $d_2$ (around 20 meters). During this phase, passes were promoted between 
players inside the situational area, which, from the network perspective, resulted in a higher clustering coefficient (Fig. \ref{fig:f03}A) and a lower shortest-path 
distance between them (Fig. \ref{fig:f03}B). In addition, trying to keep the game inside a situational area promoted the creation of short passes, reducing the risks of losing the ball, as opposed to long passes.

On the other hand, the {\it defensive phase} started as soon as the opponent had the ball and was based on the creation of a large ``interception space" to increase the chance of recovery. Similarly to the attacking phase, two regions were organized and coordinated over a radius of 20 meters. The fact that FCB played more advanced towards the opponent’s goal than their rivals, (see Fig. \ref{fig:f02}A) together with the coordinated pressure, made any eventual recovery more dangerous, increasing the probability of subsequent shooting actions.

In addition to dividing the field into various areas, a distinctive factor of FCB was the role players adopted and their area of specialization. The fact that FCB promoted generalist 
players is linked to what is known as ``total football" \cite{wilson2018}. 
In effect, with the FCB’s playing style, all players could play the ball, recover it, and score. As a consequence, a possible interpretation could be that more 
generalist profiles tend to generate more connections, take more advantage of the space, and generate different game options, leading to more 
complex passing networks. 
In addition, a team based on generalist players forces opponents to spend more energy and work harder at coordination. For example, a team without an obvious centre-forward player generates ambiguity
and uncertainty for the three or four opposing defenders. One consequence of the promotion of generalist players is the the fact that, 
from the $18$ field players that played more than 1000 minutes during the season, only 3 of them did not score a goal (Milito, Maxwell and Abidal). 

Furthermore, a number of generalist players promoted the arousal of spontaneous playing patterns, that is, different ball flows and/or positions 
 for players who are successfully scoring goals, passing, or recovering the ball. By managing the right trajectories and the right supporting 
 positions, the opponent was forced to cover more ground running and increasing his fatigue. Controlling the ball while the opponent run out 
 of energy leaded to much better positions for gaining superiority, creating surprises, and obtaining opportunities to score.

At the same time, it is worth noting the existence of a core group of players whose participation in the passing networks was higher than the rest. This fact is 
indicated by the high eigenvalue $\lambda_1$ (Fig. \ref{fig:f06}C), 
the high heterogeneity in the centrality of the players (Fig. \ref{fig:f06}E) and the existence of a player with the maximum centrality higher than the other teams (Xavi) (see Fig. \ref{fig:f06}F).
In addition, the existence of this core could be related to the fact that the algebraic connectivity is reduced when analyzing 50-pass networks
 (an indicator of the existence of groups), since an underlying core-periphery structure, combining ``leading" players with ``follower" players, 
 may lead to the existence of two identifiable communities. In this way, the existence of 4-5 players that carried and passed the ball most 
 often could be translated into the existence of a certain distributed leadership
 in the different situational areas and phases of the game – while the other players followed, coordinately, the game carried out by these leaders. 

Finally, we have to remark that the tactical organization was carefully planned and trained by Guardiola and his technical staff, and it was not a matter of serendipity \cite{wilson2018}. 
 In fact, this style of playing was one of the FCB’s signatures and it was promoted at lower categories of the team. In this way, seven out of the ten 
 players that played more than $1000$ minutes during the $2009/2010$ were raised up at {\it La Masia}, the FCB youth academy. 
 In addition, three of them (Xavi, Iniesta and Messi) were designed as the three finalists of the {\it Ballon D’Or} at the end of that season, which is given to the best football player along the whole season.
 
Summarizing, we have identified a series of particular  network properties that make Guardiola's Barcelona a team different from the rest, allowing the interpretation of the reported network parameters.
We believe that further studies taking into account the spatiotemporal evolution of football passing networks, together with recent approaches including the construction of 
network-of-networks \cite{gao2011,aguirre2013}, multilayer networks \cite{kivela2014,boccaletti2014} or hypernetworks \cite{ramos2017} could further enhance the
 understanding of how football teams, in particular, and sport teams in general, organize and evolve along a match and what are the key factors 
 that determine their performance. Furthermore, despite our results are focused on team performance, they can be adapted to evaluate single players and their
 contribution to the team. This change of ``scale", would imply some collateral issues, such as the difference in the number of matches played by each player or 
 the fact that the position a player has in the
 team unavoidably affects his/her network properties. However, we believe that this kind of new approaches will be incorporated, in the years to come, to complement
 classical metrics of player performance.

\section*{Methods}
\subsection*{Construction of the passing networks}

Datasets, provided by Opta, consists of all passes completed along a football match by each team of the Spanish national league (``La Liga") for the season $2009/2010$. Specifically, consists
of a set of $380$ matches, $38$ per team. For each pass, we have the information about: (i) the player who passes the ball, (ii) the player who receives the ball, (iii) 
the position ($x$ and $y$ coordinates) of the sender/receiver players and (iv)
the time at which the pass was made (see Tab. \ref{tab:tab01} for details).
Since we are concerned about the game of FCB, we focused on all matches played by this team, and 
analyze the passing networks of FCB and its rivals. We construct networks in two 
different ways. On one hand, we obtain the match {\it average passing networks}, where nodes are players and links represent 
the number of passes between them. Note that links are unidirectional and weighted according to the number of passes 
between players. To ease comparison between networks, each titular player is assigned a node at the beginning of the match. If 
a player is changed, the new player occupies the node of the previous player. In this way, we assure
that all networks have eleven players, focusing on the structure of the network as a whole instead of the performance of isolated players.

\begin{table}[h]
\centering
\begin{tabular}{cccccccc}
\hline

\noalign{\smallskip}
Time (seconds) & Team & Player 1 & $x_1$ & $y_1$ & Player 2 & $x_2$ & $y_2$ \\
\noalign{\smallskip}
\bottomrule[1.2pt]

\noalign{\smallskip}
... & ...  & ... & ... & ... & ... & ... & ...  \\
\noalign{\smallskip}

\noalign{\smallskip}
$355$ & F.C. Barcelona  & Busquets & 32.35 & 58.35  & Xavi & 41.20 & 61.90  \\
\noalign{\smallskip}

\noalign{\smallskip}
$359$ & F.C. Barcelona  & Xavi & 50.35 & 62.35  & Messi  & 60.70 & 64.80  \\
\noalign{\smallskip}

\noalign{\smallskip}
$363$ & F.C. Barcelona  & Messi & 70.35 & 60.55  & Henry  & 82.70 & 56.50  \\
\noalign{\smallskip}

\noalign{\smallskip}
... & ...  & ... & ... & ... & ... & ... & ...  \\
\noalign{\smallskip}

\hline
\end{tabular}
\caption{\textbf{Example of the dataset structure.} Time, in seconds, corresponds to the moment at which the pass is made. Player 1 and player 2 are, respectively,
the sender and receiver of the pass, while $x_{1,2}$ and $y_{1,2}$ are the coordinates of both players, in field units (bounded, at both axis, between $0$ and $100$).}\label{tab:tab01}
\end{table}

On the other hand, we construct the {\it ``50-pass networks"} with the aim of accounting for the temporal evolution of the game. $50$-pass 
networks contain only $50$ consecutive passes and are assigned the time of 
the last of these passes. This way, when the match begins, we wait for the first $l=50$ passes to occur and, at this moment called $t_0$, we construct and 
analyze the {\it 50-pass} network $G_{t_0}$. Next, each time a new pass is made, we disregard the oldest of the passes of the network and 
include the new one, assigning the time of the last pass $t$ to the new network $G_t$. This kind of networks has two advantages compared to the averaged
ones: (i) it accounts for the fluctuations of the network parameters along the match and (ii) it has exactly the same number of nodes and 
links for both teams, which detaches the influence of the absolute number of passes
and focuses only on the structural differences between networks. It is worth noting that the number of passes to construct the network 
could be modified to another quantity, however it should be low enough to account
to the fluctuations occurring during the match (i.e., avoiding averages) but long enough to guarantee the connectivity between 
all nodes of the network. In our case, we analyzed the effects of using different number $l$ of
passes and chose $l=50$ as a trade-off value.

\subsection*{Definition of network metrics}

\subsubsection*{Centroid coordinates and dispersion}

$\langle X \rangle$ and $\langle Y \rangle$ {\it centroid coordinates} correspond to the average position of all pases of the network, 
i.e., all passes of the match in the average network and only $50$ of them in the $50$-pass passing networks. Specifically, we 
only consider the position from where the pass is sent. Values are
given in field coordinates, which, in both axis, range from $0.00$ to $100.00$. In this way, the center of the field corresponds to 
coordinates $[50.00, 50.00]$ and the center of the opponent's goal is [100.00, 50.00] 
(being [0.00,50.00] the center of the own team's goal).
The {\it centroid dispersion} $Cent_{disp}$ corresponds to the standard deviation of the distances of the players with regard to the position of the network centroid.

\subsubsection*{Clustering coefficient}

In general, the {\it local clustering coefficient} of a node $i$ is obtained as the percentage of the nodes directly connected to it that, in turn, are connected between them. 
This measure can be averaged along the $N$ nodes of the network to obtain the {\it average clustering coefficient}. However, when the network is weighted, we can not simply account for the number of nodes connected between them but, also,
 how the link weights are distributed.
This is the case of passing networks, where the number of passes between pairs of players is not constant. 
In this way, we use the weighted clustering coefficient $C_w(i)$ to measure the likelihood that neighbours of a given player $i$ will also be connected between them \cite{ahnert2007}:

\begin{align}
& C_w(i)=\frac{\sum_{j,k}w_{ij}w_{jk}w_{ik}}{\sum_{j,k}w_{ij}w_{ik}}
\end{align}

where $j$ and $k$ are any two players of the team and $w_{ij}$ and $w_{ik}$ the number of passes between a third player $i$ and both them. Finally, 
the clustering coefficient of the whole network is obtained by
averaging $C_w(i)$ over all players, i.e., $C=\frac{1}{N}\sum^N_{i=1} C_w(i)$. Note that, the weighted version of the clustering coefficient characterizes 
the tendency of the team to form balanced triangles between players and it is a measure of local robustness.

\subsubsection*{Shortest-path length}

In a passing network, the {\it shortest path length} $d$ is the minimum number of players that must be traversed by the ball to go from one player to any other. 
Since passing networks are weighted (i.e., the number of passes between players is different), we have to take into account the different weights
 of the links, considering that, the higher the weight, the shorter the topological distance between two nodes. The topological 
 length $l_{ij}$ of the link between two players $i$ and $j$ is defined as the inverse of the link weight, $l_{ij} = 1/ w_{ij}$. However, when 
 computing $d$ for weighted networks, the shortest-path length between a pair of players may not be a direct link, since there could 
 exist a shorter path by combining two (or more) alternative links. Therefore, we compute the minimal shortest-path $p_{ij}$ 
 between all pairs of players using the Dijkstra's algorithm \cite{dijkstra1959}. Next, we define the average shortest path $d$ of the whole team as:

\begin{equation}
d=\frac{1}{N(N-1)}\sum_{i,j_{\:\:i\neq j}}p_{ij}
\end{equation}
where $N=11$ is the total number of players of the team.

\subsubsection*{Largest eigenvalue of the adjacency matrix}
The {\it largest eigenvalue} $\lambda_1$ of the weighted adjacency matrix $A$ of a network is a measure of the network strength \cite{aguirre2013}. The weighted adjacency matrix $A$ is a 
$N\times N$ matrix whose elements $a_{ij}$
contain the number of passes going from player $i$ to player $j$. The largest eigenvalue of $A$ is bounded by 
 the average number of passes between players $\langle S \rangle$, 
as $\lambda_1\geq \langle S \rangle$, and also by  $s_{max}  \geq  \lambda_1\geq max(\langle S \rangle , \sqrt{s_{max}})$ \cite{vanmieghem2011}, where $s_{max}$ is the 
maximum number of passes that a player has made to any other player of his team. As a rule of thumb, networks with higher number of links (passes) will 
have a higher $\lambda_1$ and networks with the important nodes connected between them
(known as assortative networks) will also have higher $\lambda_1$ than networks where the hubs 
(i.e., important players) are not directly connected between them. 

\subsubsection*{Algebraic Connectivity}

The {\it algebraic connectivity} $\tilde{\lambda}_2$ corresponds to the second smallest eigenvalue of the Laplacian matrix $\tilde{L}$, which is defined as $\tilde{L}=S-A$, with $A$ being the weighted adjacency matrix and  $S$ a diagonal matrix whose $i$-elements are the sum of the passes made by player $i$. 
The  algebraic connectivity  is closely related to both structural and dynamical properties of networks \cite{masuda2017,newman2010,vanmieghem2011}. 
On one hand, algebraic connectivity is an indicator of the modular structure of a network \cite{fortunato2016}: The lower the $\tilde{\lambda}_2$, the clearer 
the existence of independent groups inside the network, with the
limit value of $\tilde{\lambda}_2=0$ indicating the existence of, at least, two disconnected groups in the network. In the framework of multilayer networks, 
one can interpret the value of $\tilde{\lambda}_2$ as a way to quantify structural integration 
and segregation of different network layers \cite{radicchi2013}. On the other hand, $1/\tilde{\lambda}_2$ is proportional to the time required to reach 
equilibrium in a linear diffusion process \cite{gomez2013}. Additionally, the time $t_{sync}$ to reach synchronization of an ensemble of phase 
oscillators that are linearly and diffusively coupled is also proportional to $1/\tilde{\lambda}_2$ \cite{almendral2007}. 

\subsubsection*{Eigenvector Centrality: Maximum value and dispersion}

The {\it eigenvector centrality} $ec(i)$ of a player $i$ is a measure of node importance that is obtained by calculating the eigenvector $v_1$ 
associated to the largest eigenvalue $\lambda_1$ of the weighted adjacency matrix $A$. The eigenvector centrality is a measure of node
 importance that takes into account the number of all directed connections a player (node) has. Furthermore, two factors 
 contribute to increase the value of eigenvector centrality: (i) a higher number of direct connections to other players (note that connections are weighted) and (ii)
to be connected to other nodes that, in turn, also have a high centrality. In this way, important players are those that are (highly) connected to other important players of the team.

\subsubsection*{50-pass network time}

The {\it $50$-pass network time} $t_{diff}$ is the time required to construct a $50$-pass network. It is obtained subtracting the time of the first pass of the network from the time of the last pass. Teams with shorter
$t_{diff}$ are those that generate more passes in less time.

\subsection*{Statistical analysis} 
All parameters of Figs. \ref{fig:f02} and \ref{fig:f03} have been compared pairwise with the Wilcoxon ranksum test, as the number of observations to compare 
was small enough to prevent us from safely assuming normality ($< 40$ in most of the cases). However, the t-statistics yielded the same rejections of the
 null hypothesis (central tendency equality, median or mean) in all parameters. As the number of comparisons (20) would raise type I errors, p-values 
 have been corrected for multiple comparisons with non parametric false discovery rate (FDR) \cite{benjamini2001} for $\alpha = 0.01$, which changed some results on 
 the verge on significance. After FDR correction, $\alpha = 2.7132e^{-04}$ ($0.004$ if we originally set alpha at $0.01$). $<Y>$ was not statistically 
 significant for any threshold, which is expected, and should not change with the number of observations. On the contrary, eigenvector 
 centrality dispersion ($p = 0.0042$) should be considered significantly different only if we keep alpha unaltered. After FDR correction we cannot
  state any difference confidently. The centroid dispersion remains significant in any case, although just barely ($p = 2.6e^{-4}$). 
  In both cases, conclusions must be taken with caution, and we would need more statistical power (i.e., more data) to assert confidently that there are statistically significant differences in those parameters. 
  


\section*{Acknowledgements}

Authors acknowledge Johann Mart\'inez, Jos\'e Luis Herrera Diestra, Roberto L\'opez, Esteban Granero, Chechu Fern\'andez Conde and Joan Bennassar for fruitful conversations. J.M.B. is founded by MINECO, Spain (FIS2017-84151-P). Datasets were provided by Opta.

\section*{Additional information}
{\bf Competing Interests:} The authors declare no competing interests.

\section*{Author contributions statement}
J.M.B., J.B. and F.S. conceived the ideas behind the paper. J.M.B. conducted the analysis of the datasets. I.E. carried out the statistical analysis. J.M.B., J.B., I.E. and F.S. analyzed the results. All authors wrote and reviewed the manuscript. 


\end{document}